\documentclass{article}
\usepackage{txfonts,epsfig,graphicx,natbib,url,hyperref,twoopt}

\bibpunct{(}{)}{;}{a}{}{,}    %% natbib format choice 

\renewcommandtwoopt{\cite}[3][][]{\citealp[#1][#2]{#3}}    
\newcommandtwoopt{\citepads}[3][][]{\href{http://adsabs.harvard.edu/abs/#3}%
                                        {\citep[#1][#2]{#3}}}
\newcommandtwoopt{\citetads}[3][][]{\href{http://adsabs.harvard.edu/abs/#3}%
                                        {\citet[#1][#2]{#3}}}
\newcommandtwoopt{\citeads}[3][][]{\href{http://adsabs.harvard.edu/abs/#3}%
                                        {\cite[#1][#2]{#3}}}
\newcommandtwoopt{\citeyearads}[3][][]{\href{http://adsabs.harvard.edu/abs/#3}%
                                        {\citeyear[#1][#2]{#3}}}

\newcount\longrefs
\def\aap{\ifnum\longrefs=1 {Astron.\ Astrophys.}\else 
                           {A\hbox{\rm \&}A}\fi}
\def\aapr{\ifnum\longrefs=1 {Astron.\ Astrophys.\ Rev.}\else 
                            {A\hbox{\rm \&}AR}\fi}
\def\aaps{\ifnum\longrefs=1 {Astron.\ Astrophys.\ Suppl.}\else 
                            {A\hbox{\rm \&}A Suppl.}\fi}
\def\aj{\ifnum\longrefs=1 {Astron.\ J.}\else 
                          {AJ}\fi} 
\def\ao{\ifnum\longrefs=1 {Applied Optics}\else 
                           {Appl.\ Opt.}\fi} 
\def\aspcs{\ifnum\longrefs=1 {Astron.\ Soc.\ Pacific Conf. Series}\else 
                           {ASP Conf.\ Ser.}\fi} 
\def\apj{\ifnum\longrefs=1 {Astrophys.\ J.}\else 
                           {ApJ}\fi} 
\def\apjl{\ifnum\longrefs=1 {Astrophys.\ J. Lett.}\else 
                            {ApJ}\fi} 
\def\aplett{\ifnum\longrefs=1 {Astrophys.\ J. Lett.}\else 
                            {ApJ}\fi} 
\def\apjs{\ifnum\longrefs=1 {Astrophys.\ J. Suppl.}\else 
                            {ApJS}\fi}
\def\apss{\ifnum\longrefs=1 {Astrophys.\ and Space Science}\else 
                            {Astrophys.\ Space Sci.}\fi}
\def\araa{\ifnum\longrefs=1 {Ann.\ Rev.\ Astron.\ Astrophys.}\else 
                            {ARA\hbox{\rm \&}A}\fi}
\def\azh{\ifnum\longrefs=1 {Astronomicheskii Zhurnal}\else 
                            {Astron.\ Zhur.}\fi}
\def\baas{\ifnum\longrefs=1 {Bull.\ Am.\ Astron.\ Soc.}\else 
                            {BAAS}\fi}
\def\bain{\ifnum\longrefs=1 {Bull.\ Astronom.\ Institutes Netherlands}\else
                            {Bull.\ Astr.\ Inst.\ Neth.}\fi}
\def\gca{\ifnum\longrefs=1 {Geochim.\ Cosmochim.\ Acta}\else 
                           {Geochim.\ Cosmochim.\ Acta}\fi}
\def\grl{\ifnum\longrefs=1 {Geophys.\ Res.\ Lett.}\else 
                           {Geoph.\ Res.\ Lett.}\fi}
\def\iaucirc{\ifnum\longrefs=1 {IAU Circulars}\else 
                          {IAU Circ.}\fi}
\def\ip{\ifnum\longrefs=1 {in press}\else 
                          {in press}\fi}
\def\jgr{\ifnum\longrefs=1 {J.\ Geophys.\ Res.}\else 
                           {J.\ Geophys.\ Res.}\fi}  
\def\jrasc{\ifnum\longrefs=1 {J.\ Royal Astron.\ Soc.\ Canada}\else 
                           {JRAS Can.}\fi}  
\def\mnras{\ifnum\longrefs=1 {Mon.\ Not.\ Roy.\ Astron.\ Soc.}\else 
                             {MNRAS}\fi} 
\def\nat{\ifnum\longrefs=1 {Nature}\else 
                           {Nat}\fi}
\def\pasj{\ifnum\longrefs=1 {Pub.\ Astron.\ Soc.\ Japan}\else 
                            {PASJ}\fi} 
\def\pasp{\ifnum\longrefs=1 {Pub.\ Astron.\ Soc.\ Pacific}\else 
                            {PASP}\fi} 
\def\physscr{\ifnum\longrefs=1 {Physica Scripta}\else 
                            {Phys.\ Scrip.}\fi} 
\def\planss{\ifnum\longrefs=1 {Planetary \& Space Science}\else 
                            {Plan. \& Space Sci.}\fi} 
\def\procspie{\ifnum\longrefs=1 {Proc.\ SPIE}\else 
                            {Proc.\ SPIE}\fi} 
\def\qjras{\ifnum\longrefs=1 {Quarterly J.\ Royal Astron.\ Soc.}\else 
                            {QJRAS}\fi} 
\def\sa{\ifnum\longrefs=1 {Soviet Astron..}\else 
                               {Sov.\ Astron.}\fi}
\def\skytel{\ifnum\longrefs=1 {Sky \& Telescope}\else 
                            {Sky \& Tel.}\fi} 
\def\solphys{\ifnum\longrefs=1 {Solar Phys.}\else 
                               {Sol.\ Phys.}\fi}
\def\ssr{\ifnum\longrefs=1 {Space Science Rev.}\else 
                               {Space\ Sci.\ Rev.}\fi}

\def\linkadscode#1{http://adsabs.harvard.edu/abs/#1}

\newlength{\skipblanklength}
 
\def\dutch{\def\refname{Referenties}\def\abstractname{Samenvatting}%
  \def\bibname{Bibliografie}\def\chaptername{Hoofdstuk}%
  \def\appendixname{Bijlage}\def\contentsname{Inhoudsopgave}%
  \def\listfigurename{Lijst van figuren}%
  \def\listtablename{Lijst van tabellen}%
  \def\indexname{Index}\def\figurename{Figuur}\def\tablename{Tabel}%
  \def\partname{Deel}\def\enclname{Bijlage(n)}\def\ccname{Ter attentie van}%
  \def\headtoname{Aan}\def\headpagename{Pagina}%
  \def\today{\number\day\space\ifcase\month\or januari\or februari\or%
     maart\or%
     april\or mei\or juni\or juli\or augustus\or september\or oktober\or%
     november\or december\fi \space\number\year}%
  \typeout{
              >>>>> use hlatex209 for Dutch hyphenation <<<<< 
         }}

\DeclareFontFamily{OT1}{mvs}{}
\DeclareFontShape{OT1}{mvs}{m}{n}{<-> fmvr8x}{}

\newcounter{onefig} \newcounter{fignumber}
\newcount\nocaptions \newcount\nofigures \newcount\figwidth
\newcount\viewgraphs \newcount\printlabel
\long\def\skipfigure #1\viewout{}   %% use \skipfigure....\viewout
  \def\paper{}  \def\figlabel{} 
\long\def\nextfig#1{\setcounter{figure}{\value{fignumber}}
  \addtocounter{fignumber}{1}
  \ifnum \viewgraphs=1 \pagestyle{empty} \fi 
  \ifnum\value{onefig}=0 #1 \fi                 
  \ifnum\value{onefig}=\value{fignumber} #1 \fi}
\def\figwidths#1#2{\ifnum \nocaptions=1 #2mm \else #1mm \fi}  
\def\picplace{\framebox[80mm]{\rule{0cm}{1cm}}}
\def\paper#1{}  %% redefine for separate-figure identification line
\long\def\plotfig#1#2{\ifnum \nofigures=1 \picplace \else #2 \fi}
\long\def\captiontext#1{\ifnum \nofigures=1 \raggedright \fi 
   \ifnum \nocaptions=1 \paper
     \ifnum \viewgraphs=0 
       \newline  \mbox{}\hrulefill\mbox{} \newline 
       \ifnum \printlabel=1 \{{\em \figlabel}\}\newline \fi
     \fi 
   \else \ifnum \printlabel=1 \{{\em \figlabel}\}\newline \fi
     #1 \fi}

\newcount\panelwidth \newcount\panelheight 
\newcount\bxmin \newcount\bymin \newcount\bxmax \newcount\bymax
\newcount\tbxmin \newcount\tbymin
\newcount\tpanelwidth \newcount\tpanelheight \newcount\tpdif
\def\panelsize #1,#2;{\panelwidth=#1 \panelheight=#2}  
\def\setbb #1,#2;#3,#4;#5,#6;{% UNITS: bp (from ghostview)
  \tbxmin=#1 \tbymin=#2    %% full box (axis titles) lower left corner
  \bxmin=#3 \bymin=#4      %% bare box (ticks only) lower left corner
  \bxmax=#5 \bymax=#6}     %% upper right corner
\def\barepanel #1{%
  \ifnum\panelheight=0 
    \tpdif=\bymax \advance\tpdif by -\bymin
    \multiply \tpdif by \panelwidth
    \tpanelheight=\tpdif
    \tpdif=\bxmax \advance\tpdif by -\bxmin
    \divide \tpanelheight by \tpdif
  \else \tpanelheight=\panelheight \fi
  \ifnum\panelwidth=0 
    \tpdif=\bxmax \advance\tpdif by -\bxmin
    \multiply \tpdif by \panelheight
    \tpanelwidth=\tpdif
    \tpdif=\bymax \advance\tpdif by -\bymin
    \divide \tpanelwidth by \tpdif
  \else \tpanelwidth=\panelwidth \fi
  \epsfig{file=#1,silent=,%
     bbllx=\bxmin bp,bblly=\bymin bp,bburx=\bxmax bp,bbury=\bymax bp,clip=,%
     width=\tpanelwidth mm,height=\tpanelheight mm}}
\def\labelypanel #1{% TeX permits only integer arithmetic, so bp and mm
  \ifnum\panelheight=0 
    \tpdif=\bymax \advance\tpdif by -\bymin
    \multiply \tpdif by \panelwidth
    \tpanelheight=\tpdif
    \tpdif=\bxmax \advance\tpdif by -\bxmin
    \divide \tpanelheight by \tpdif
  \else \tpanelheight=\panelheight \fi
  \ifnum\panelwidth=0 
    \tpdif=\bxmax \advance\tpdif by -\bxmin
    \multiply \tpdif by \panelheight
    \tpanelwidth=\tpdif
    \tpdif=\bymax \advance\tpdif by -\bymin
    \divide \tpanelwidth by \tpdif
  \else \tpanelwidth=\panelwidth \fi
  \tpdif=\bxmax \advance\tpdif by -\tbxmin
  \multiply \tpanelwidth by \tpdif
  \tpdif=\bxmax \advance\tpdif by -\bxmin
  \divide \tpanelwidth by \tpdif
  \epsfig{file=#1,silent=,%
    bbllx=\tbxmin bp,bblly=\bymin bp,bburx=\bxmax bp,bbury=\bymax bp,%
    clip=,width=\tpanelwidth mm,height=\tpanelheight mm}}
\def\labelxpanel #1{%
  \ifnum\panelheight=0 
    \tpdif=\bymax \advance\tpdif by -\bymin
    \multiply \tpdif by \panelwidth
    \tpanelheight=\tpdif
    \tpdif=\bxmax \advance\tpdif by -\bxmin
    \divide \tpanelheight by \tpdif
  \else \tpanelheight=\panelheight \fi
  \ifnum\panelwidth=0 
    \tpdif=\bxmax \advance\tpdif by -\bxmin
    \multiply \tpdif by \panelheight
    \tpanelwidth=\tpdif
    \tpdif=\bymax \advance\tpdif by -\bymin
    \divide \tpanelwidth by \tpdif
  \else \tpanelwidth=\panelwidth \fi
  \tpdif=\bymax \advance\tpdif by -\tbymin
  \multiply \tpanelheight by \tpdif
  \tpdif=\bymax \advance\tpdif by -\bymin
  \divide \tpanelheight by \tpdif
  \epsfig{file=#1,silent=,%
    bbllx=\bxmin bp,bblly=\tbymin bp,bburx=\bxmax bp,bbury=\bymax bp,%
    clip=,width=\tpanelwidth mm,height=\tpanelheight mm}}
\def\labelxypanel #1{%
  \ifnum\panelheight=0 
    \tpdif=\bymax \advance\tpdif by -\bymin
    \multiply \tpdif by \panelwidth
    \tpanelheight=\tpdif
    \tpdif=\bxmax \advance\tpdif by -\bxmin
    \divide \tpanelheight by \tpdif
  \else \tpanelheight=\panelheight \fi
  \ifnum\panelwidth=0 
    \tpdif=\bxmax \advance\tpdif by -\bxmin
    \multiply \tpdif by \panelheight
    \tpanelwidth=\tpdif
    \tpdif=\bymax \advance\tpdif by -\bymin
    \divide \tpanelwidth by \tpdif
  \else \tpanelwidth=\panelwidth \fi
  \tpdif=\bxmax \advance\tpdif by -\tbxmin
  \multiply \tpanelwidth by \tpdif
  \tpdif=\bxmax \advance\tpdif by -\bxmin
  \divide \tpanelwidth by \tpdif 
  \tpdif=\bymax \advance\tpdif by -\tbymin 
  \multiply \tpanelheight by \tpdif
  \tpdif=\bymax \advance\tpdif by -\bymin
  \divide \tpanelheight by \tpdif
  \epsfig{file=#1,silent=,%
    bbllx=\tbxmin bp,bblly=\tbymin bp,bburx=\bxmax bp,bbury=\bymax bp,%
    clip=,width=\tpanelwidth mm,height=\tpanelheight mm}}

\def\CC{\par \vspace*{-2ex} \footnotesize \baselineskip=8pt \begin{verbatim}}

\long\def\startignore #1\stopignore{}   %% use \startignore....\stopignore

\def\setlistparams{         
  \topsep=0.7ex                 %% ADAPT: parskip=0: 0.7;  parskip=1: -1.2ex
  \itemsep=0.7ex                %% space between items
  \leftmargini=3ex}             %% dashes at beginning of line 
\setlistparams                  %% recall after type changes 

\newcounter{alistindex}       %% problems: a)  b) etc

\newcounter{romenumnr}

\newlength{\minipagewidth}

\newsavebox{\boxcontent}
\newcommand{\ovalhead}[1]{
  \unitlength=1cm
  \sbox{\boxcontent}{\mbox{~~{#1}~~}}
  \begin{center}
    \ifdim\wd\boxcontent>6ex 
    \ifdim\wd\boxcontent<8cm 
    \begin{picture}(8,3) \thicklines     
      \put(4.0,0.8){\oval(8,1.6)} 
      \put(0.0,0.7){\parbox{8cm}{
         \begin{center} \usebox{\boxcontent} \end{center}}}
    \end{picture}
    \else \ifdim\wd\boxcontent<12cm 
    \begin{picture}(12,3) \thicklines     
        \put(6.0,0.8){\oval(12,1.6)} 
        \put(0.0,0.7){\parbox{12cm}{
           \begin{center} \usebox{\boxcontent} \end{center}}}
    \end{picture}
    \else
    \begin{picture}(16,3) \thicklines     
        \put(8.0,0.8){\oval(16,1.6)} 
        \put(0.0,0.7){\parbox{16cm}{
           \begin{center} \usebox{\boxcontent} \end{center}}}
    \end{picture}
    \fi \fi \fi
  \end{center}} 

\newcounter{headnr}            
\newcounter{subheadnr}[headnr]
\newcounter{subsubheadnr}[subheadnr]
\font\dropfont= cmr12 scaled \magstep5
\def\dropcap#1#2{{\noindent
    \setbox0\hbox{\dropfont #1}\setbox1\hbox{#2}\setbox2\hbox{(}%
    \count0=\ht0\advance\count0 by\dp0\count1\baselineskip
    \advance\count0 by-\ht1\advance\count0by\ht2
    \dimen1=.5ex\advance\count0by\dimen1\divide\count0 by\count1
    \advance\count0 by1\dimen0\wd0
    \advance\dimen0 by.25em\dimen1=\ht0\advance\dimen1 by-\ht1
    \global\hangindent\dimen0\global\hangafter-\count0
    \hskip-\dimen0\setbox0\hbox to\dimen0{\raise-\dimen1\box0\hss}%
    \dp0=0in\ht0=0in\box0}#2}

\def\rmit#1{{\it #1}}              %% italics (RR style, Kluwer)
\def\etal{\rmit{et al.}}           %% use \etal\ for space behind it        
           
\def\ie{\rmit{i.e.,}}              %% , required (Webster 1681)
\def\eg{\rmit{e.g.,}}              %% , required (Webster 1681)
\def\cf{cf.}                       %% no Latin, always Roman (Webster 1686)

\def\specchar#1{\uppercase{#1}}    %% to be redefined for A&A, small caps
\def\CaII{\mbox{Ca\,\specchar{ii}}}

\def\Halpha{\mbox{H\hspace{0.1ex}$\alpha$}} %% \Halpha\ for space behind it

\def\Hbeta{\mbox{H\hspace{0.2ex}$\beta$}}

\def\NaD{\mbox{Na\,\specchar{i}\,D}}    %% use \NaD\ for space behind it

\def\CaIIK{\mbox{Ca\,\specchar{ii}\,\,K}}       %% use \CaIIK\ for space
\def\CaIIH{\mbox{Ca\,\specchar{ii}\,\,H}}

\def\HK{\mbox{H\,\&\,K}}
\def\KtwoV{\mbox{K$_{2V}$}}

\def\HtwoV{\mbox{H$_{2V}$}}

\def\level #1 #2#3#4{$#1 \: ^{#2} \mbox{#3} ^{#4}$}   

\def\rmb{{\rm b}}              %% use for indices etc. 

\def\kms{\hbox{km$\;$s$^{-1}$}}

\def\is{\!=\!}                             %% = in text for tighter spacing
\def\kf{($k_h,f$)}                         %% f - k_h
\def\={\hbox{$\!=\!$}}                     %% less space around =

\def\mathstacksym#1#2#3#4#5{\def#1{\mathrel{\hbox to 0pt{\lower 
    #5\hbox{#3}\hss} \raise #4\hbox{#2}}}}

\mathstacksym\lta{$<$}{$\sim$}{1.5pt}{3.5pt} % less than approximately
\mathstacksym\gta{$>$}{$\sim$}{1.5pt}{3.5pt} % greater than approximately
\mathstacksym\lrarrow{$\leftarrow$}{$\rightarrow$}{2pt}{1pt} % equilibrium
\mathstacksym\lessgreat{$>$}{$<$}{3pt}{3pt} %% less or greater

\def\rmit#1{{\rm #1}}   %% italics out - Kluwer went Roman?

\newcommand{\captionfonts}{\it}
\makeatletter  % Allow the use of @ in command names
\long\def\@makecaption#1#2{%
  \vskip\abovecaptionskip
  \sbox\@tempboxa{{\captionfonts #1: #2}}%
  \ifdim \wd\@tempboxa >\hsize
    {\captionfonts #1: #2\par}
  \else
    \hbox to\hsize{\hfil\box\@tempboxa\hfil}%
  \fi
  \vskip\belowcaptionskip}
\makeatother   % Cancel the effect of \makeatletter

\begin{document}
\pagestyle{headings}

\title{%% \mbox{} \vspace{-5ex} 
    Waves in the chromosphere: observations\footnote{%
    Appeared in the proceedings of a NATO Advanced Research Workshop
    held at Budapest in 2002 edited by
    \citetads{2003twis.book.....E} %% Budapest procs
    but not entered or even listed on ADS through laxness of the
    publisher and editors.  Submitted to arXiv Astro-ph on December 6,
    2010 to remedy this neglect and to demonstrate my newest latex
    trick: in on-screen reading, clicking on the year in each citation
    opens the corresponding ADS abstract page in your browser.
    Explanation: \url{http://www.astro.uu.nl/\~rutten}.}}

\author{R.J. Rutten\\[1ex]
        Sterrekundig Instituut, Utrecht University\\
        Institute for Theoretical Astrophysics, Oslo University}

\date{January 23, 2003}

%%\vspace{-10ex}
\maketitle

%%%%%%%%%%%%%%%%%%%%%%%%%%%%%%%%%%%%%%%%%%%%%%%%%%%%%%%%%%%%%%%%%%%%%%%%%%%%
\begin{abstract}

  I review the literature on observational aspects of waves in the
  solar chromosphere in the first part of this contribution.
  High-frequency waves are invoked to build elaborate cool-star
  chromosphere heating theories but have not been detected decisively
  so far, neither as magnetic modes in network elements nor as
  acoustic modes in below-the-canopy internetwork regions.
  Three-minute upward-propagating acoustic shocks are thoroughly
  established through numerical simulation as the cause of
  intermittent bright internetwork grains, but their pistoning and
  their role in the low-chromosphere energy budget remain in debate.
  Three-minute wave interaction with magnetic canopies is a newer
  interest, presently progressing through numerical simulation.
  Three-minute umbral flashes and running penumbral waves seem a
  similar acoustic-shock phenomenon awaiting numerical simulation.
  The low-frequency network Doppler modulation remains enigmatic.

  In the second part, I address low-frequency ultraviolet brightness
  variations of the internetwork chromosphere in more detail.  They
  contribute about half of the internetwork brightness modulation and
  presumably figure in cool-star basal flux.  They appear to be a
  mixture of inverse-contrast granular overshoot at small scales and
  gravity-wave interference at mesogranular scales.  I present TRACE
  evidence for the latter interpretation, and speculate that the
  low-frequency brightness minima map canopy heights.
\end{abstract}
%%%%%%%%%%%%%%%%%%%%%%%%%%%%%%%%%%%%%%%%%%%%%%%%%%%%%%%%%%%%%%%%%%%%%%%%%%%%

%%%%%%%%%%%%%%%%%%%%%%%%%%%%%%%%%%%%%%%%%%%%%%%%%%%%%%%%%%%%%%%%%%%%%%%%%%%%
%%%%%%%%%%%%%%%%%%%%%%%%%%%%%%%%%%%%%%%%%%%%%%%%%%%%%%%%%%%%%%%%%%%%%%%%%%%%
\section{Introduction}  \label{sec:introduction}
%%%%%%%%%%%%%%%%%%%%%%%%%%%%%%%%%%%%%%%%%%%%%%%%%%%%%%%%%%%%%%%%%%%%%%%%%%%%
%%%%%%%%%%%%%%%%%%%%%%%%%%%%%%%%%%%%%%%%%%%%%%%%%%%%%%%%%%%%%%%%%%%%%%%%%%%%

Waves in the solar chromosphere may be divided in many different ways:
  high versus low chromosphere,
  below the canopy versus above the canopy,
  network versus internetwork,
  acoustic versus magnetic versus gravity mode,
  linear versus shocks,
  quiet versus active region,
  standing versus propagating versus evanescent,
  global versus local,
  short-period versus three-minute versus five-minute versus long-period,
  longitudinal versus transverse versus torsional.
And, in particular, whether they heat appreciably or negligibly.  The
latter issue is the most frequently quoted motivation for
chromospheric wave studies and for their claims to fame, but it seems
to me that identifying the nature of the observed modulations and the
underlying processes and interactions should come first -- and,
personally, I regard the physics as more interesting than the
eventually resulting overall energy budget, a mere summation at the
very end with the answer already known as balancing photon losses.

%%%%%%%%%%%%%%%%%%%%%%%%%%%%%%%%%%%%%%%%%%%%%%%%%%%%%%%%%%%%%%%%%%%%%%%%%%%%
\section{Overview}   \label{sec:overview}
%%%%%%%%%%%%%%%%%%%%%%%%%%%%%%%%%%%%%%%%%%%%%%%%%%%%%%%%%%%%%%%%%%%%%%%%%%%%

In this section I briefly review the literature\footnote{Note added
  December 5, 2010: All ADS abstracts for almost all solar physicists
  cited in this paper are collected at
  \url{http://www.astro.uu.nl/\~rutten/solar_abstracts}.}  using the
above splits as guideline.

%%%%%%%%%%%%%%%%%%%%%%%%%%%%%%%%%%%%%%%%%%%%%%%%%%%%%%%%%%%%%%%%%%%%%%%%%%%%
\paragraph{Chromosphere.}
%%%%%%%%%%%%%%%%%%%%%%%%%%%%%%%%%%%%%%%%%%%%%%%%%%%%%%%%%%%%%%%%%%%%%%%%%%%%

Even this term is non-trivial.  In the standard one-dimensional
modeling of Vernazza, Avrett \& Loeser
(\citeyearads{1973ApJ...184..605V}, % VALI
\citeyearads{1976ApJS...30....1V}, % VALII
\citeyearads{1981ApJS...45..635V}) % VALIII
and \citetads{1993ApJ...406..319F}, % FALC
it denotes the plane-parallel layers between the temperature minimum
and the onset of the coronal temperature rise.  Within this
definition, UV continua shortward of 1600\,\AA\ come from the
chromosphere (VALIII Fig.~36 in \citeads{1981ApJS...45..635V}), and
strong optical lines are also chromospheric -- but not always.  For
example, the \NaD\ lines reach $\tau \is 1$ above the VALIII
temperature minimum, but they are photospheric in their intensity
response and do not map the chromospheric temperature rise in their
VALIII emergent profiles
(\citeads{1992A&A...265..237B}). %T AA NLTE formation NaI and KI
Even \CaIIK\ filtergrams primarily sample the upper photosphere since
the intensity-encoding thermal creation (rather than the
velocity-encoding last scattering) of most of the observed photons
takes place well below the VALIII minimum.  In addition, the existence
of a ubiquitous temperature starting at $h=500$~km has been put into
doubt, first by Ayres' CO modeling (\eg\
\citeads{1981ApJ...245.1124A}; %? ApJ FTS CO I
\citeads{1981ApJ...244.1064A}; %? thermal bifurcation
\citeads{1986ApJ...304..542A}; %? ApJ CO + CaIIK
\citeads{1989ApJ...338.1033A}; %? ApJ CO formation
\citeads{1990ApJ...363..705A}; %? ApJ time resolved
\citeads{1996ApJ...460.1042A}) %T Ayres+Rabin, CO revisited
and subsequently by the Carlsson--Stein acoustic shock modeling (see
below).

Hence, I prefer to define ``chromosphere'' -- harking back to its
eclipse origin = the purple color of \Halpha\ plus \Hbeta\ off-limb
emission -- as the solar regime characteristically sampled by Balmer
lines as a complex mass of fibrils, and taking the latter to represent
(incomplete!) mappings of magnetic canopies\footnote{The
  definition implies that chromospheric studies must include \Halpha\
  even though its formation is singularly awkward through mixing
  thickness and thinness with NLTE opacity and source function
  sensitivities including Zanstra-like ionisation-plus-recombination
  photon conversions.  An unpleasant but inescapable conclusion to one
  prefering clean lines such as \CaII\ \HK.  Quantitative \Halpha\
  mapping requires considerable work on reliable \Halpha\
  interpretation, comparable to the ongoing efforts in Stokes profile
  inversion.}.
%

%%%%%%%%%%%%%%%%%%%%%%%%%%%%%%%%%%%%%%%%%%%%%%%%%%%%%%%%%%%%%%%%%%%%%%%%%%%%
\paragraph{Canopies.}
%%%%%%%%%%%%%%%%%%%%%%%%%%%%%%%%%%%%%%%%%%%%%%%%%%%%%%%%%%%%%%%%%%%%%%%%%%%%

The concept comes from
\citetads{1980SoPh...68...49G}, %? Giovanelli, canopies
\citetads{1982SoPh...79..247J}, %? Jones+Giovanelli, response to canopies
\citetads{1982SoPh...79..267G} %? Giovanelli+Jones, low canopies everywhere
and is portrayed in older models and recent simulations (\eg\
\citeads{1993A&A...268..736B}; %T Bunte+Solank+Steiner, wine glass model
\citeads{2002ApJ...564..508R}; %T Rosenthal++, IN reflections
\citeads{2002AN....323..196B}) %T Bogdan Potsdam Thinkshop, mix+interfere
as smoothly upward spreading field hovering dome-like over essentially
field-free internetwork.  This is a simplification.  If we define the
chromosphere to start at canopy height (the height where the plasma
beta drops through unity), its lower boundary will actually be a very
warped surface, offset by dynamical flows and with large topological
variations defined by the small-scale and large-scale field strength
and connectivity, as partially delineated by \Halpha\ as short local
and long distant fibril connections (\cf\
\citeads{2002SoPh..207..223S}). %C Schrijver+Title, network topology

%%%%%%%%%%%%%%%%%%%%%%%%%%%%%%%%%%%%%%%%%%%%%%%%%%%%%%%%%%%%%%%%%%%%%%%%%%%%
\paragraph{High versus low chromosphere.}
%%%%%%%%%%%%%%%%%%%%%%%%%%%%%%%%%%%%%%%%%%%%%%%%%%%%%%%%%%%%%%%%%%%%%%%%%%%%

This distinction now means well above and just above the canopy,
respectively.  Most of the references in this review pertain to the
low chromosphere or even the upper photosphere (say $h=400-800$~km).
Higher up, most of the recent wave literature employs CDS and SUMER
spectrometry to discuss to what height the internetwork three-minute
oscillations penetrate (\eg\
\citeads{1997ASPC..118..284S}; %T Aspe I SUMER+CDS
\citeads{1997ApJ...486L..63C}; %? Carlsson+Judge+Wilhelm, SUMER internetwork
\citeads{1997ApJ...490L.195J}; %? Judge+Carlsson+Wilhelm, SUMER network
\citeads{1998SoPh..181...51D}; % Doyle++ CDS
\citeads{1998ApJ...503L..95C}; %T Curdt + Heinzel, SUMER Lyman lines
\citeads{1999SoPh..184..253G}; %? Gouttebroze++, SUMER oscillations
\citeads{1999A&A...347..335D}; %? Doyle+vdOord+OShea+Banerjee, SUMER
\citeads{2000ApJ...531.1150W}; %C Wikstol+Hansteen+Carlsson+Judge SUMER
\citeads{2001ApJ...554..424J}; %T Judge+Tarbell+Wilhelm, shadows
\citeads{2001ApJ...561..420M}). %C McIntosh+Judge, shadows
Such penetration is likely to vary strongly with the actual field
geometry.  Oscillation analyses in tandem with moss and coronal loop
diagnostics are yet scarce (but see Ineke de Moortel's contribution in
this volume).

Theoretical insight comes from beautiful Oslo simulations
(\citeads{2002ApJ...564..508R}; %T Rosenthal++, IN reflections
\cf\
\citeads{2002AN....323..196B}). %T Bogdan Potsdam Thinkshop, mix+interfere
I wonder whether such simulations for different stellar parameters
may, at long last, explain the Wilson-Bappu relation between stellar
luminosity and \CaII\ \HK\ peak width
(\citeads{1957ApJ...125..661W}).  % Wilson-Bappu
It must describe fluxtube atmosphere properties, rather than
non-magnetic atmospheric stratifications (\eg\
\citeads{1979ApJ...228..509A}; %? Wilson-Bappu scaling laws
\citeads{1983A&A...128..311K}), %? Kneer, Wilson-Bappu from CO cooling
since the emission peaks come from the magnetic component.

%%%%%%%%%%%%%%%%%%%%%%%%%%%%%%%%%%%%%%%%%%%%%%%%%%%%%%%%%%%%%%%%%%%%%%%%%%%%
\paragraph{Network versus internetwork.} 
%%%%%%%%%%%%%%%%%%%%%%%%%%%%%%%%%%%%%%%%%%%%%%%%%%%%%%%%%%%%%%%%%%%%%%%%%%%%

This division is also overly simplistic.  The verdict on internetwork
fields isn't yet in, but they exist undoubtedly at some level of field
strength and scale of spatial organization.  \CaIIK\ and TRACE UV
image sequences show extended zones of enhanced brightness (with
respect to the darkest ``cell centers'') as aureoles around network
(\cf\ Fig.~\ref{fig:canopy}).  Acoustic maps show similar three-minute
power aureoles when sampling power below the canopy and power shadows
above it (\eg\
\citeads{1992ApJ...392..739B}; %T Braun++, local acoustics
\citeads{1992ApJ...394L..65B}; % Brown+Bogdan+Lites+Thomas
\citeads{1993ApJ...415..847T}; % Toner+LaBonte power maps
\citeads{1998ApJ...504.1029H}; %T Hindman+Brown, 3-min halos
\citeads{1999ApJ...510..494L}; %? Lindsey+Braun, acoustic moats + halos
\citeads{1999ApJ...513L..79B}; %T Braun+Lindsey, halos
\citeads{2000ApJ...537.1086T}; %T Thomas+Stanchfield, halos
\citeads{2001ApJ...548L.237M}; %? Macintosh etal network mode conversion
\citeads{2001ApJ...554..424J}; %T Judge+Tarbell+Wilhelm, shadows
\citeads{2001ApJ...561..420M}; %C McIntosh+Judge, shadows
\citeads{2001A&A...379.1052K}). %T TRACE1

In addition, \CaIIK\ and TRACE UV image sequences also show evidence
of ``magnetic flashers'', presumably isolated fluxtubes on their way
to or from the network concentrations (Brandt \etal,
\citeyearads{1992ASPC...26..161B}, % Brandt+Rutten+Shine+Trujillo
\citeyearads{1994ssm..work..251B}; % Brandt+Rutten+Shine+Trujillo
\citeads{1998SoPh..179..253N}; %T Nindos+Zirin, K2v grains & IN fields
\citeads{1999ApJ...517.1013L}; % Lites+Rutten+Berger 2v grains
\citeads{2001A&A...379.1052K}). %T TRACE1
Their isolation may be helpful in trying to identify chromospheric
tube modes excited by convective buffeting without amplitude loss from
phase mixing over multiple fluxtubes.

%%%%%%%%%%%%%%%%%%%%%%%%%%%%%%%%%%%%%%%%%%%%%%%%%%%%%%%%%%%%%%%%%%%%%%%%%%%%
\paragraph{Acoustic versus magnetic versus gravity modes.}
%%%%%%%%%%%%%%%%%%%%%%%%%%%%%%%%%%%%%%%%%%%%%%%%%%%%%%%%%%%%%%%%%%%%%%%%%%%%

Or mixtures, of course.  {\em Magnetic modes\/} are often invoked to
convey energy from photospheric buffeting up along network fluxtubes
(\eg\
\citeads{1999ApJ...519..899H}; %T Hasan+Kalkofen, granule fluxtube excitation
\citeads{2000ApJ...535L..67H}) %T Hasan+Kalk+AvB, buffeting excitation
but have not been convincingly diagnosed yet, except negatively as
suppression of acoustic wave power by conversion at canopies
(\citeads{2001ApJ...548L.237M}; %? McIntosh++, network mode conversion
\citeads{2001ApJ...561..420M}; %C McIntosh+Judge, shadows
\citeads{2002ApJ...564..508R}). %T Rosenthal++, IN reflections

{\em Internal gravity waves\/} should be ``copiously excited'' in
granular overshoot according to theory (\eg\
\citeads{1963ApJ...137..914W}; %T Whitaker: coronal heating grav waves
Lighthill %? IAU Symp 28 429
in \citeads{1967IAUS...28.....T}; % Thomas Aerodynamic
\citeads{1967SoPh....2..385S}; %? Stein
\citeads{1977SoPh...54..269S}; %T Schmieder linear theory
Mihalas \& Toomre
\citeyearads{1981ApJ...249..349M}, %T Mihalas+Toomre I
\citeyearads{1982ApJ...263..386M}), %T Mihalas+Toomre II
but they are difficult to detect, being small-scale and propagating
slantedly.  The observational evidence is mostly indirect
(\citeads{1968ApJ...152..557F}; %T Frazier: also grav waves
\citeads{1976SoPh...47..435S}; % Schmieder
\citeads{1978A&A....70..345C}; % Cram power+phase Abel transform diagrams
\citeads{1980ApJ...236L.169B}; %? Brown+Harrison: from brightness
\citeads{1981A&A....95..221D}; %? Durrant+Nesis
\citeads{1984MmSAI..55..147S}; %? Staiger++: grav waves
\citeads{1987A&A...175..263S}; %? Staiger: also grav waves
\citeads{1989A&A...213..423D}; % Deubner+Fleck I
\citeads{1991A&A...244..492B}; %T Bonet etal KI grav waves
\citeads{1991A&A...252..827K}), %T Komm+Mattig+Nesis
with the clearest demonstration coming from wavenumber- and
frequency-resolved \kf\ phase-difference spectra
(\citeads{1993A&A...274..584K}; %T Kneer+Uexkull, K grains, resonances, grav
\citeads{1997A&A...324..704S}; %% Straus+Bonaccini, mesoscale
\citeads{2001A&A...379.1052K}). %T TRACE1
Whether they affect upper-photosphere or low-chromosphere energy
balances is not known.  See also Section~\ref{sec:background} below.

%%%%%%%%%%%%%%%%%%%%%%%%%%%%%%%%%%%%%%%%%%%%%%%%%%%%%%%%%%%%%%%%%%%%%%%%%%%%
\paragraph{Linear versus shock behavior.}
%%%%%%%%%%%%%%%%%%%%%%%%%%%%%%%%%%%%%%%%%%%%%%%%%%%%%%%%%%%%%%%%%%%%%%%%%%%%

At least in the case of the three-minute oscillation, protests against
large non-linearity
(\citeads{1991mcch.conf....6D}) % Deubner Heidelberg
were silenced by the successful reproduction of so-called \CaII\
\HtwoV\ grain behavior in the celebrated simulations of Carlsson and
Stein
(\citeyearads{1994chdy.conf...47C}, %T C+S Oslo Miniworkshop H2v simulation
\citeyearads{1996ASPC..109..119C}, %? Carlsson+Stein, K2v CSW Florence
\citeyearads{1997ApJ...481..500C}; %C Carlsson+Stein, K2v grain simulation
\citeads{1997ASSL..225..261S}) %T Stein+Carlsson, SCORe review
of the spectral time sequences of
\citetads{1993ApJ...414..345L}. %T Lites+Rutten+Kalkofen, network dynamics
The Carlsson-Stein reproduction of complex spectral \CaIIH\ core
evolution patterns identified the three-minute oscillation beyond
doubt with upward propagating acoustic shock trains, as proposed
earlier by \citetads{1970SoPh...11..347A}, % Athay HK asymmetries
\citetads{1972SoPh...22..375C}, % Cram multi-component models
\citetads{1974SoPh...38..109L}, % Liu+Skumanich
\citetads{1987A&A...177..283M}, % Mein+Mein+Malherbe+Dame
\citetads{1982ApJ...258..393L}, % Leibacher+Gouttebroze+Stein
Rutten \& Uitenbroek
(\citeyearads{1991mcch.conf...48R}, % Rutten+Uitenbroek Heidelberg
\citeyearads{1991SoPh..134...15R}) % Rutten+Uitenbroek k2v grains
and \citetads{1992A&A...253..586R}. % Rammacher+Ulmschneider
However, the nature of the photospheric pistoning remains in debate
(\citeads{1998SoPh..179..253N}; %? Nindos+Zirin, K2v grains & IN fields
\citeads{1999ApJ...517.1013L}; % Lites+Rutten+Berger H2v grains nonB
\citeads{1999ApJ...523..450W}; %T Worden+Harvey+Shine, TRACE-mg, nonB
\citeads{2000A&A...363..279S}; %T Sivaraman++, revisit K2v grains
\citeads{2000ApJ...541..468S}; %C Skartlien+Stein+Nordlund, waves collapsar
\citeads{2002A&A...390..681H}, % Hoekzema+Rimmele+Rutten
as does the role or absence of a role in chromospheric heating
(\citeads{1995ApJ...440L..29C}; %C Carlsson+Stein, does chrom T-rise exist?
\citeads{1997A&A...324..587T}; %T Theurer+Ulmschneider+Kalkofen, 3-min peak
\citeads{1999ApJ...521L.141K}; %T Kalkofen++, chrom T-rise from high f
\citeads{2001ApJ...557..376K}). %? Kalkofen: case against dark chrom
See also Section~\ref{sec:background} below.

%%%%%%%%%%%%%%%%%%%%%%%%%%%%%%%%%%%%%%%%%%%%%%%%%%%%%%%%%%%%%%%%%%%%%%%%%%%%
\paragraph{Active region oscillations.}
%%%%%%%%%%%%%%%%%%%%%%%%%%%%%%%%%%%%%%%%%%%%%%%%%%%%%%%%%%%%%%%%%%%%%%%%%%%%

A direct active-region counterpart to the quiet-sun three-minute
oscillation consists of the three-minute oscillations producing {\em
  umbral flashes\/}
(\citeads{1969SoPh....7..351B}). %T Beckers+Tallant, discovery flashes
They show even more outspoken nonlinear character (\eg\
\citeads{1984ApJ...285..368T}; %T Thomas+Cram+Nye, flash obs
\citeads{1986ApJ...301.1005L}), %T Lites, CaIIH+He10830
and are most likely similar nearly-acoustic shock trains running up
along the radial fields above umbrae as suggested by Lites
(\citeyearads{1992sto..work..261L}, % Lites review Cambridge NATO workshop
\citeyearads{1994chdy.conf....1L}) % Lites review Oslo miniworkshop
Some are observed to penetrate up into coronal loops in UV and EUV
data (\eg\ Brynildsen \etal,
\citeyearads{1999ApJ...511L.121B}, %T Brynildsen++, SUMER 3-min TR oscs
\citeyearads{2002SoPh..207..259B}; %T Brynildsen++, SOHO+TRACE spot osc
Maltby \etal,
\citeyearads{1999SoPh..190..437M}, %T Maltby++, CDS+SUMER spot oscs up to TR
\citeyearads{2001A&A...373L...1M}; %T Maltby++, osc only in umbral plume
\citeads{1999SoPh..187..261S}) %T Schrijver++, TRACE new view
and in microwave observations
(\citeads{1999SoPh..185..177G}; %T Gelfreikh++, microwave umbra osc
\citeads{2001ApJ...550.1113S}; %T Shibasaki, microwave umbral oscs
\citeads{2002A&A...386..658N}). %T Nindos++, VLA TR base modulation 25 km

Similarly, {\em running penumbral waves\/}
(\citeads{1972ApJ...178L..85Z}; %T Zirin+A.Stein, discovery rpw
\citeads{1992SoPh..138...93A}; %T Alissandrakis++, rpw in Halpha
\citeads{1997ApJ...478..814B}; %T Brisken+Zirin, rpw
Christopoulou \etal,
\citeyearads{2000A&A...354..305C}, %T C+G+K, rpw's, no clear relation umbra
\citeyearads{2000A&A...363..306G}, %T C+G+K, phot sunspot wavs in+out
\citeyearads{2001A&A...375..617C}, %T C+G+K, rpw's
\citeyearads{2002ApJ...576..561G}) %T C+Muglach+G, TRACE rpw's in+out
may harbor complex wave reflection and mode conversion as postulated
by
\citeads{2002AN....323..196B} %T Bogdan Potsdam Thinkshop, mix+interfere
from comparison to thinner fluxtubes, but recent high-resolution data
from three solar telescopes on La Palma suggest strongly that running
penumbral waves and umbral flashes are both upward-propagating
acoustic shock trains, seen differently through difference in field
alignment with the line of sight and with the apparent horizontal
spreading primarily a mapping of the field geometry
(\citeads{2003A&A...403..277R}). %T La Palma umbral flashes
Oslo simulations as those of
\citetads{1997ApJ...481..500C} %C Carlsson+Stein, K2v grain simulation
are needed for decisive diagnosis, including ``calibration'' of the
polarimetric umbral-flash inversions of
\citetads{2001ApJ...550.1102S} %T Socas+Trujillo+Ruiz, umbral flashes
through less indirect forward modeling.

The existence and nature of oscillations in the sunspot magnetic field
itself remain contested (\eg\
\citeads{1998ApJ...497..464L}; %T Lites++ umbral B oscillations
\citeads{1999SoPh..187..389B}; %CT Balthasar power rings
\citeads{2002AN....323..317S}: % Staude Potsdam Thinkshop
\citeads{2000SoPh..191...97K}), %T Kupke+Labonte+Mickey: spot oscs
with
\citetads{2002A&A...392.1095S} %T Settele+Sigwarth+Muglach: spot oscs
warning against instrumental crosstalk and initial results from
infrared spectropolarimetry indicating that opacity variations are an
alternative explanation (\cf\
\citeads{2002AN....323..254C}; % Collados Potsdam Thinkshop, IR sunspots
\citeads{2003ApJ...588..606K}). % Khomenko+Collados+Bellot
\citetads{2002ApJ...576..561G} %T C+Muglach+G, TRACE rpw's in+out
found indications from TRACE data that running penumbral waves even
spread into sunspot moats.

Chromospheric oscillations above plage have not received much
attention in the literature, but Rita Ryutova's presentation of
beautiful TRACE 171\,\AA\ time slices from Richard Shine show
interesting braiding with shorter braid period at larger magnetic
filling factor (see her contribution in these proceedings).

%%%%%%%%%%%%%%%%%%%%%%%%%%%%%%%%%%%%%%%%%%%%%%%%%%%%%%%%%%%%%%%%%%%%%%%%%%%%
\paragraph{Standing versus propagation.} 
%%%%%%%%%%%%%%%%%%%%%%%%%%%%%%%%%%%%%%%%%%%%%%%%%%%%%%%%%%%%%%%%%%%%%%%%%%%%

This issue is also debated in the three-minute oscillation literature
(Deubner \etal,
\citeyearads{1992A&A...266..560D}, %? Deubner+Fleck+Schmitz+Straus
\citeyearads{1996A&A...307..936D}; %T Deubner+Waldschik+Steffens, cavity NaD
\citeads{1995A&A...302..277S}; %T Steffens+Hofmann+Deubner+Fleck, 6 mHz mode
\citeads{1996A&A...308..192H}; %T Hofmann+Steffens+Deubner, phase K2v
\citeads{1998IAUS..185..427D}). % Deubner Kyoto review phases
with simulations showing that at least some of the observed
standing-wave behavior comes from reflections off the grain-forming
shocks themselves and off canopies
(\citeads{1999ASPC..184..206C}; % Carlsson+Stein ASPE Potsdam
\citeads{2002ApJ...564..508R}). %T Rosenthal++, IN reflections

Another standing versus propagation issue concerns the nature of the
so-called pseudo-ridges above the cutoff frequency in \kf\ diagrams,
also a global-versus-local issue.
\citetads{1994ApJ...428..827K} %T Kumar, properties acoustic sources
described them as interference between directly emitted and
once-bounced outgoing waves (that propagate up rather than being
evanescent), using ``interference'' as a misleading mathematical term
describing low-$l$ Fourier decomposition.  Physically there is no
actual wave interference, but simply power addition in \kf\ diagrams
at those \kf\ locations that sample one-bounce horizontal spatial
wavelengths at the corresponding temporal frequency, just another
expression of the single-bounce three-minute power ridge in
time-distance plots
(\citeads{1993Natur.362..430D}) % Duvall++ time-distance
and exhibiting the Duvall dispersion law
(\citeads{1982Natur.300..242D}; %? Duvall law, see ARAA 22 605
Eq.~2.14 of \citeads{1984ARA&A..22..593D}).  Below the cutoff
frequency, the acoustic ridges describe evanescent $p$-modes, with
small phase delays compared to the photosphere governed by
non-adiabaticity
  (\citeads{2001A&A...379.1052K}). %T TRACE1

%%%%%%%%%%%%%%%%%%%%%%%%%%%%%%%%%%%%%%%%%%%%%%%%%%%%%%%%%%%%%%%%%%%%%%%%%%%%
\paragraph{Short-period versus three-minute versus 
five-minute versus low-frequency modulation.}
%%%%%%%%%%%%%%%%%%%%%%%%%%%%%%%%%%%%%%%%%%%%%%%%%%%%%%%%%%%%%%%%%%%%%%%%%%%%

These terms are all often misnomers.  By {\em short-period waves\/}
one simply implies those that might after all heat the chromosphere,
with Peter Ulmschneider with coworkers as tenacious champion.  After
giving up on five-minute heating when the $p$-modes were identified
and on three-minute heating when the Carlsson--Stein simulation
refuted a low-lying chromospheric temperature rise, his quest turned
to higher-frequency components not present in the Carlsson-Stein
piston (\eg\
\citeads{1995A&A...294..241S}; %T Sutmann+Ulmschneider, II = nonlinear
Theurer \etal,
\citeyearads{1997A&A...324..587T}, %T Theurer+Ulmschneider+Kalkofen, 3-min peak
\citeyearads{1997A&A...324..717T}; %T Theurer+Ulmschneider+Cuntz, freq spectra
\citeads{1999ApJ...521L.141K}) %T Kalkofen++, chrom T-rise from high f
with obvious stellar overtones
(\citeads{1996A&A...315..212U}, %T Ulmschneider+Theurer+Musielak, star fluxes
\citeads{1999ApJ...522.1053C}), %? Cuntz++, stellar CaIIK modeling
and more recently to adding longitudinal, transverse and torsional
tube waves (Ulmschneider \etal,
\citeyearads{2001A&A...374..662U}, %T Ulm++, longitudinal tube waves
\citeyearads{2001ApJ...559L.167U}; %T Ulm++, tube wave heating dwarfs
Musielak and Ulmschneider,
\citeyearads{2002ApJ...573..418M}, %T Musielak+Ulm+Rosner, fluxtube waves IV
\citeyearads{2002A&A...386..606M}, %T Musielak+Ulm, transverse tube waves II
\citeyearads{2002A&A...386..615M}) %T Musielak+Ulm, transverse tube waves III
as well as detailed theoretical prediction of the mix of acoustic and
tube waves that should explain observed cool-star chromospheric photon
losses with the magnetic filling factor as activity scaling parameter
(Fawzy \etal,
\citeyearads{2002A&A...386..971F}, %T Fawzy++, acoustic+magnetic heating I
\citeyearads{2002A&A...386..983F}, %T Fawzy++, acoustic+magnetic heating II
\citeyearads{2002A&A...386..994F}). %T Fawzy++, acoustic+magnetic heating III

Heating by 10--100~mHz waves, outside and inside fluxtubes, is
probable; the question is to what extent.  Observationally, they are
hard to see since they require fast cadence and high angular
resolution and suffer response function loss by vertical wavelengths
fitting in contribution functions (\eg\
\citeads{1980A&A....84...99S}; %? Mein+Mein, HK vel response
\citeads{1980A&A....91..251D}) %? Durrant, vel response
and through large sensitivity to seeing noise (\eg\
\citeads{1983A&A...121..291E}; %% Endler+Deubner, seeing
\citeads{1984MmSAI..55..135D}; % Deubner+Endler+Staiger phase and seeing
\citeads{1994ssm..work..159L}). % Lites+Rutten+Thomas Soesterberg
Indeed, the trials to detect high-frequency waves remain inconclusive
(\eg\ \citeads{1976A&A....51..189D}; %? short-period power modulation
\citeads{1979ApJ...231..570L}; % Lies+Chipman oscillation phase delays
\citeads{1981A&A....97..310M}; %T NMein+Schmieder, wave flux III
\citeads{1990A&A...228..506D}; % Deubner+Fleck III, cell-interior
\citeads{2001A&A...379.1052K}). %T TRACE1

The chromospheric {\em three-minute\/} and {\em five-minute}
oscillations are discussed above.  The names imply broad-band
frequency domains, not specific periodicities but just a shift in
dominance from five-minute to three-minute periodicity when rising
from the photosphere up to the chromosphere (in the internetwork at
least).  It is therefore also wrong to call them chromospheric and
photospheric, respectively.  The original Leighton-Simon-Noyes era
diagram in Noyes
(\citeyearads{1967IAUS...28..293N}), % Noyes review Aerodynamic Phenomena
reprinted in Rutten
(\citeyearads{1995ESASP.376a.151R}, % Rutten Asilomar
\citeyearads{2001ASPC..223..117R}) % Rutten CSW Tenerife, dynamics review
clearly demonstrates the gradual change.

Finally, {\em low-frequency oscillations\/} is a misnomer when the
observed modulation is not oscillatory. This may hold for the observed
low-frequency Dopplershift power of chomospheric network which may
result from convective fluxtube buffeting (\cf\ Kneer and von
Uexk{\"u}ll
\citeyearads{1985A&A...144..443K}, % Kneer+vonUexkull Halpha oscs
\citeyearads{1986A&A...155..178K}; % Kneer+vonUexkull Halpha oscs
\citeads{1989A&A...208..290V}; %T Uxkull++, network heating
\citeads{1993ApJ...414..345L}; %T Lites+Rutten+Kalkofen, network dynamics
\citeads{2000ApJ...535L..67H}), %T Hasan+Kalk+AvB, buffeting excitation
and also for low-frequency low-chromosphere internetwork oscillations
if granular overshoot rather than internal gravity waves dominates the
observed low-frequency power.

%%%%%%%%%%%%%%%%%%%%%%%%%%%%%%%%%%%%%%%%%%%%%%%%%%%%%%%%%%%%%%%%%%%%%%%%%%%%
\section{Low-frequency internetwork background modulation}  
\label{sec:background}
%%%%%%%%%%%%%%%%%%%%%%%%%%%%%%%%%%%%%%%%%%%%%%%%%%%%%%%%%%%%%%%%%%%%%%%%%%%%

I devote the remainder of this contribution to the topic on which I
concentrated in my oral presentation at the meeting: the nature of the
low-frequency background in low-chromosphere (``upper photosphere'' in
the under-the-canopy definition) image sequences.  I have long been
puzzled by the mesh-like background pattern underlying \KtwoV\ grains
in \CaIIK\ image sequences (\cf\
\citeads{1999ApJ...517.1013L}). % Lites+Rutten+Berger H2v grains
The upshot is that I believe the answer to be a mixture of granular
overshoot plus gravity-wave interference at slightly larger scales.  I
advocate this interpretation with selected results from TRACE data
analyzed by J.M.~Krijger in his PhD thesis
(\citeads{Krijger-thesis2002}). %RR no ADS
The diagrams in this section also illustrate various points made
above\footnote{%
  Note added on December 5, 2010: part of this work was published in
  \citetads{2003A&A...407..735R} % Rutten+Krijger evidence g waves
  but not all diagrams and ideas given here survived the referee.}.

The observation and reduction details are given in
\citetads{2001A&A...379.1052K}. %T TRACE1
The displays below come from the very quiet-sun data taken on October
14, 1998, when TRACE registered image sequences in its 1700, 1600 and
1550\,\AA\ ultraviolet passbands and also in white light.  The
combination permits comparison of photospheric brightness patterns to
the co-spatial and co-temporal ultraviolet ones.  Such comparisons
used to be made with groundbased telescopes combining \CaIIK\ and
continuum image registration (\eg\ Hoekzema \etal,
\citeyearads{Hoekzema+Rutten+Brandt+Shine1998}, %T A&A, WORM
\citeyearads{Hoekzema+Rutten1998}, %T A&A, GBP = G & K
\citeyearads{Hoekzema+Rimmele+Rutten2002}) %T FLUXCA
but TRACE furnishes better quality thanks to the absence of seeing in
space\footnote{But the tide turns, now that speckle reconstruction,
  phase-diverse restoration and adaptive optics also do away with
  (most of) the seeing, permitting higher angular resolution than
  TRACE's 1~arcsec.  Much sharper images now result \eg\ from our
  Dutch Open Telescope which is presently being equipped with
  multi-wavelength tomography capability
  (\url{http://dot.astro.uu.nl}).}.

%%%%%%%%%%%%%%%%%%%%%%%%%%%%%%%%%%%%%%%%%%%%%%%%%%%%%%%%%%%%%%%%%%%%%%%%%%%%
\paragraph{Space-time representations.}
%%%%%%%%%%%%%%%%%%%%%%%%%%%%%%%%%%%%%%%%%%%%%%%%%%%%%%%%%%%%%%%%%%%%%%%%%%%%

Figure~\ref{fig:imslices} defines the topic.  It compares the
white-light low-photosphere scene with the co-spatial and co-temporal
1700\,\AA\ high-photosphere scene in the upper panels.  The lower
panels compare evolutionary characteristics between the two regimes.
The displays sample only very small parts of the full white-light and
1700\,\AA\ data cubes\footnote{These
  cutouts are small in order to obtain sufficient magnification.  They
  represent a very limited rendering of the full data cubes.  A better
  view of the dynamical ultraviolet scene is gained by inspecting the
  TRACE movies available at
  \url{http://www.astro.uu.nl/~rutten/trace1}.}.

%% {fig:imslices} 
%============================================================================
\begin{figure}
  \centerline{\includegraphics[width=\textwidth]{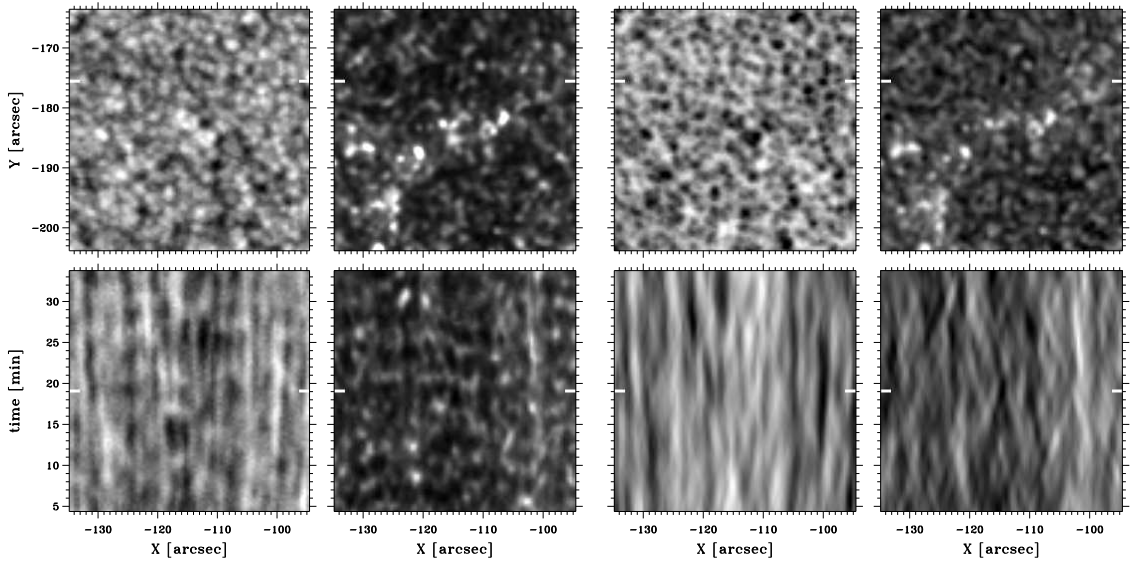}}
  \caption[]{TRACE image and time-slice samples.  First column:
    white-light intensity.  Second column: co-temporal and co-spatial
    1700\,\AA\ intensity.  The images in the upper row are small
    cut-outs of the full 256$\times$256~arcsec$^2$ field shown in
    Fig.~\ref{fig:canopy}. The time of observation is halfway the time
    slices in the lower row, as indicated by white markers.  The
    slices show the temporal intensity evolution during 30~min of
    observation, for a horizontal cut through an internetwork area
    indicated by the white markers in the upper panels.  It passes
    through weak network at the right. In the first two columns the
    greyscale is linear for all four panels, but it has been clipped
    at half the actual maximum for the 1700\,\AA\ image to enhance
    internetwork.  Third and fourth column: the same, but low-pass
    filtered (subsonic horizontal propagation only).  Third column:
    white-light brightness on a sign-reversed greyscale.  Fourth
    column: logarithm of the 1700\,\AA\ brughtness temperature.
  } \label{fig:imslices}
\end{figure}
%============================================================================

The difference between dark internetwork and bright network is obvious
in the ultraviolet image (second column).  The network grains stand
out even though their brightness is cut in half by the display
scaling.  TRACE's resolution (1~arcsec) is insufficient to resolve the
corresponding white-light ``network bright points'' (magnetic
elements) residing within the underlying intergranular lanes.

The white-light time slice in the lower-left panel displays primarily
granular evolution, with some larger-scale five-minute $p$-mode
modulation.  The ultraviolet time slice shows the dynamical behavior
characteristic of internetwork in the form of ubiquitous short-lived
three-minute oscillation sequences.  The brightest phases are called
internetwork grains.  The weak network grain near $X=-100$~arcsec
stands out by its relative longevity.  Note that the cut selection
favours internetwork; stronger network grains produce much brighter
vertical streaks.

The two righthand columns show the same data after Fourier filtering
and with modified greyscaling.  In the third column (white light)
low-pass ``subsonic' filtering, \ie\ applying a 3D Fourier ``cone''
filter to the transformed data cube which passes all signals with
apparent horizontal speed below the sound speed, has removed the
photospheric five-minute oscillation so that only the granular
evolution patterns remain.  In addition, the greyscale is
sign-reversed to simulate ``reversed granulation'' with reversed
contrast.  The resulting mesh pattern in the reversed image
illustrates the topological difference between granules and
intergranular lanes.  The corresponding time slice shows intergranular
lanes as rather long-lived bright streaks.

In the fourth column (1700\,\AA) the low-pass filtering has removed
the chromospheric three-minute oscillation and therefore emphasizes
the slower background evolution.  The background streaks are
relatively short and show larger horizontal displacements (tilts) than
the intergranular streaks.  At this resolution, there is no obvious
correspondence between the reversed low-frequency granular evolution
pattern in the third slice and the low-frequency internetwork
background modulation pattern in the fourth slice.

%%%%%%%%%%%%%%%%%%%%%%%%%%%%%%%%%%%%%%%%%%%%%%%%%%%%%%%%%%%%%%%%%%%%%%%%%%%%
\paragraph{Time-delay scatter representations.}
%%%%%%%%%%%%%%%%%%%%%%%%%%%%%%%%%%%%%%%%%%%%%%%%%%%%%%%%%%%%%%%%%%%%%%%%%%%%

Figures~\ref{fig:scatter1}--\ref{fig:scatter2} are pixel-by-pixel
dual-image correlation plots in an informative format initiated by
\citetads{1994PhDT.......347S}. % Strous thesis
Each plot is measured from a $256\times256$~px$^2$ TRACE subfield
after removal of Fourier taper edges, image conversion into brightness
temperature, and $1.5\times1.5$~arcsec$^2$ boxcar smoothing to
suppress noise.

For each pixel in the subfield, \ie\ each solar location, the
brightness temperature at one moment in the one type of image (say
white light = WL) is taken as $x$ quantity plotted horizontally, its
value at another (later) moment in the other type of image (say
1700\,\AA\ = UV) as $y$ quantity plotted vertically.  The pair defines
one point in a scatter plot.  Such pairwise comparisons are made for
all pixels (solar locations) in the whole subfield or in some
selective part of it, and repeated for 50 consecutive image pairs 
to gain high significance (millions of spatio-temporal samples).
Plot saturation (total blackness) is avoided by plotting sample
density contours instead of the individual pixel-by-pixel samples.

%% {fig:scatter1}
%===========================================================================
\begin{figure}
  \centerline{\includegraphics[width=\textwidth]{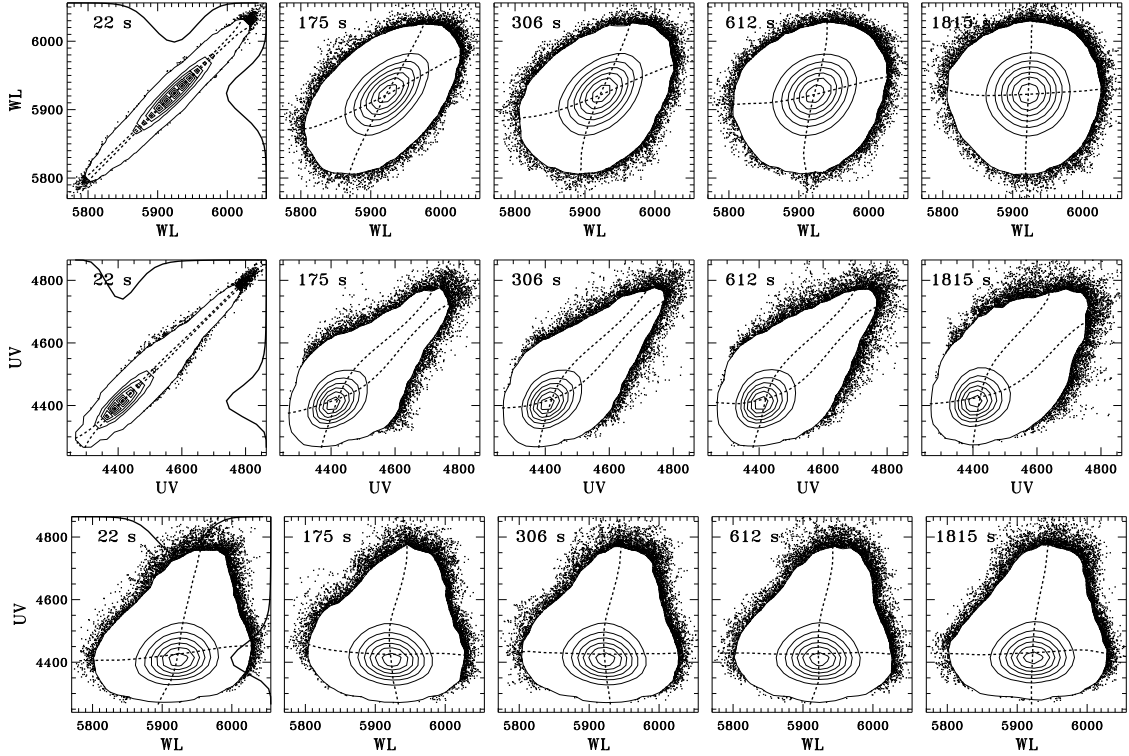}}
  \caption[]{ Strous-format scatter diagrams.  The crowded central
    parts are plotted as sample density contours to avoid plot
    saturation.  First row: time-delayed white-light brightness
    temperature in K WL$(t+\Delta t)$ against WL$(t)$.  Second row:
    time-delayed 1700\,\AA\ brightness temperature UV$(t+\Delta t)$
    against UV$(t)$.  Third row: UV$(t+\Delta t)$ against WL$(t)$.
    The time delays $\Delta t$ are specified in each panel and
    increase from left to right.  The solid curves in the first panels
    show the occurrence distributions of the quantities plotted along
    $x$ and $y$ on inverted normalized scales. The dashed curves show
    the first moments of the sample density along horizontal and
    vertical cuts through the contours.  }\label{fig:scatter1}
\end{figure}
%===========================================================================

%% {fig:scatter2}
%===========================================================================
\begin{figure}
  \centerline{\includegraphics[width=\textwidth]{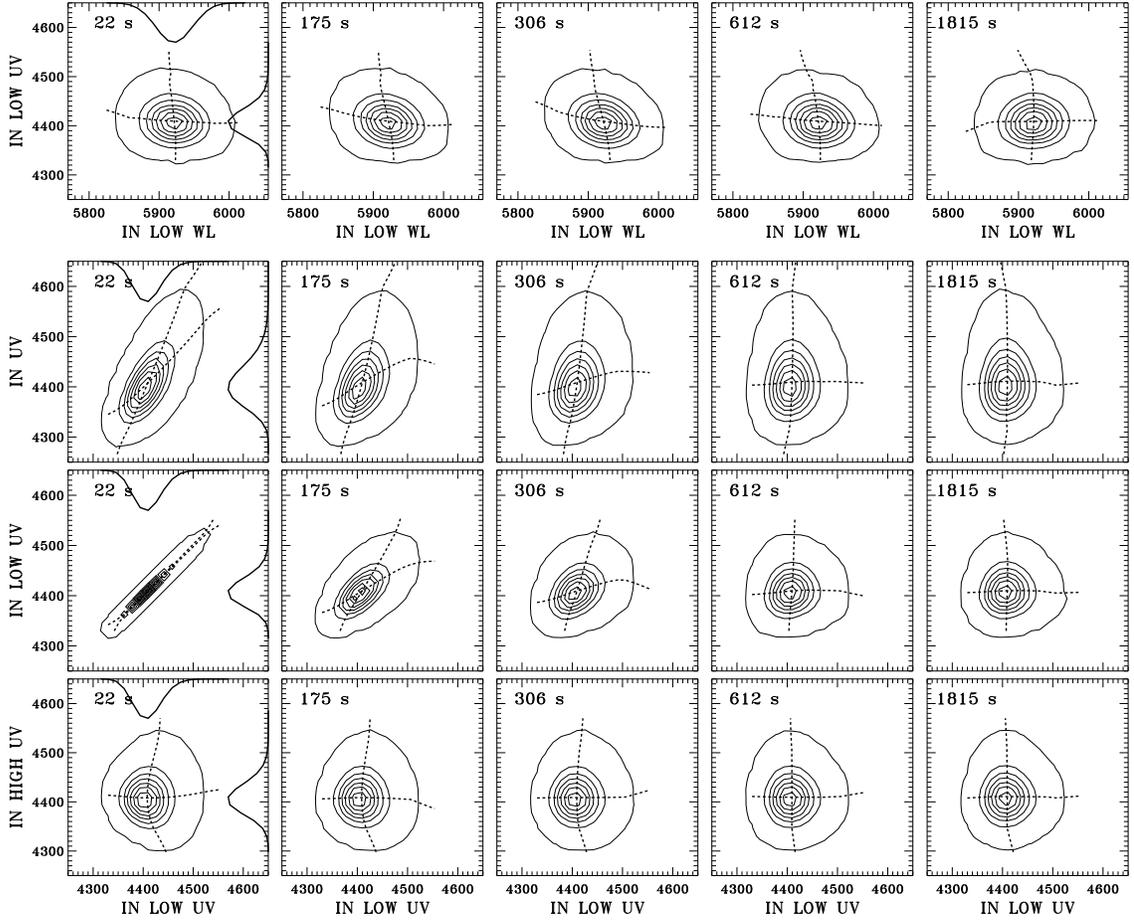}}
  \caption[]{ Time-delay scatter diagrams as in
    Fig.~\ref{fig:scatter1} but for internetwork (IN) only.  First
    row: time-delayed low-pass filtered 1700\,\AA\ brightness
    temperature UV$(t+\Delta t)$ against low-pass filtered white-light
    brightness temperature WL$(t)$.  Second row: unfiltered UV against
    low-pass UV.  Third row: low-pass UV against low-pass UV.  Fourth
    row: high-pass UV against low-pass UV, with the mean UV
    internetwork value ($T_\rmb = 4418$~K) added to the high-pass UV
    modulation for $y$-axis compatibility.  }\label{fig:scatter2}
\end{figure}
%===========================================================================

The WL--WL comparison in the first row of Fig.~\ref{fig:scatter1}
illustrates the format.  Brightness distribution curves are added in
the first panel.  They are virtually the same for the other panels and
also along $x$ and $y$ in this auto-correlation sequence.  For 100\%
correlation all samples and both first-moment (center of gravity)
curves lie along the forward diagonal.  For 100\% anticorrelation they
lie along the backward diagonal.  At the absence of any correlation
the first-moment curves become perpendicular, parallel to the axes,
and the contours become circular if the distribution function is
symmetric.  From left to right the panels have increasing time delay
between the sampling of each type of image per solar location.  The
initial panel for $\Delta t = 22$~s shows very high pattern
correlation because the granulation has not changed much during this
brief interval.  The final panel for $\Delta t = 30$~min shows absence
of pattern correlation in the form of non-aligned first-moment curves
and a nearly round bull's-eye contour pattern.  The sequence
illustrates that granulation largely loses its pattern identity over
ten minutes, completely in half an hour.

The UV--UV autocorrelation sequence in the second row illustrates the
two-component dichotomy between chromospheric network and
internetwork.  The bright upward distribution tail made up by network
grains persists over long delays, whereas the darker internetwork
(lower-left contour mountain) shows faster pattern change.

The UV--WL cross-correlation sequence in the bottom row of
Fig.~\ref{fig:scatter1} shows some persistent bright-bright
correlation due to network, slight anticorrelation for low WL at
$\Delta t = 3$~min, and subsequent lack of persistent correlation at
low UV brightness.

In Fig.~\ref{fig:scatter2} the ultraviolet signal is decomposed in
constituents by using selective data subsets for similar scatter
diagrams.  All four rows are limited to internetwork (IN) areas only.
Fourier cone filtering as in Fig.~\ref{fig:imslices} is applied to
show low-pass signals (apparent horizontal propagation below 6~\kms)
or high-pass signals (above 7~\kms) only, selecting acoustic and
non-acoustic modulation.  The corresponding reduction in sampling
statistics is offset by extending the pixel-by-pixel sampling to 120
consecutive image pairs per comparison (in all scatter plots the outer
contour lies at 100~samples per bin with 25~bins per axis).  The outer
scatter clouds are now excluded to reduce the plot file size.

The first row of Fig.~\ref{fig:scatter2} plots low-pass UV against
low-pass WL.  There is a slight but persistent anticorrelation over
time delay $\Delta t = 2-6$~min.  I attribute it to reversed
granulation and expect it to become stronger in higher-resolution data
that resolve granulation better than TRACE does.  The slightness of
the correlation shows that at the somewhat larger meso-scales imaged
properly by TRACE, something else is present that does not obey
point-to-point correlation or counter-correlation with the underlying
photosphere at any time delay.

%RR if only reverse granulation even TRACE should show high correlation
%RR unless Thijs' coregistration is bad - but then phase shouldn't be
%RR cleanly negative

The remaining three rows decompose the UV signal in high-frequency and
low-frequency components, as contributed by the latter (low-pass UV as
$x$-axis quantity).  The second row plots total UV against low-pass UV
in internetwork.  The high initial correlation, with a roughly 2:1
slope, shows that peaks in internetwork UV brightness, \ie\
internetwork grains, occur preferentially when also the slow-changing
internetwork background is bright.  This strong correlation is conform
the finding of \citetads{1983ApJ...272..355C} % Cram+Dame
that bright \CaII\ \HtwoV\ grains are invariably part of a
larger-scale modulation pattern.  The same happens in the ultraviolet
continua.  Thus, the slowly evolving background pattern and the
three-minute oscillation combine constructively to produce bright
internetwork grains, each contributing about half of the excess
brightness temperature.  This makes the background a much more
important grain co-localizer\footnote{The low-frequency background was
  not part of the Carlsson-Stein simulation.  The corresponding
  Dopplershifts have less power than the brightness modulation and
  were not part of the Carlsson-Stein iron-blend piston because they
  were removed in spectrograph drift correction by
  \citetads{1993ApJ...414..345L}.} %T Lites+Rutten+Kalkofen, network dynamics
than \eg\ acoustic events (\cf\
\citeads{2002A&A...390..681H}). % Hoekzema+Rimmele+Rutten

The next row shows a low-pass internetwork-only UV autocorrelation
sequence.  It illustrates the five-minute lifetime of the
low-frequency pattern at a given spatial position.  Horizontal
propagation -- seen as tilts in the low-pass UV time slice in the
final panel of Fig.~\ref{fig:imslices} when having an $x$ component --
causes loss of scatter correlation when the travel exceeds the
1.5~arcsec boxcar smoothing.

The final row correlates high-pass UV with low-pass UV in
internetwork.  The curved shape of the vertical moment curves
indicates that large three-minute excursions, both to high and to low
UV brightness temperature with respect to the mean after removal of
all slow variations, tend partially toward correlation with large UV
background amplitude.  This implies that the three-minute oscillation
partially has modulatory character, gaining larger oscillation
amplitude at larger background amplitude.  The effect is not large.
In addition, the contours have slightly asymmetric shape indicating
non-linearity in wave behavior.  When plotted as intensity, the
scatter clouds stretch much further upwards due to the non-linear
Planck function response in the ultraviolet.  The conversion to
brightness temperature circularizes the contours considerably.  

The conclusions from Figs.~\ref{fig:scatter1}--\ref{fig:scatter2} are,
first, that the slowly-evolving internetwork ultraviolet background
contributes about half of internetwork grain brightness excesses, in
addition to the acoustic modulation modeled by Carlsson \& Stein, and,
second, that at TRACE's meso-granular resolution this background seems
to be something else than reversed granulation.

%% {fig:Fourier}
%===========================================================================
\begin{figure}
  \centerline{\includegraphics[width=\textwidth]{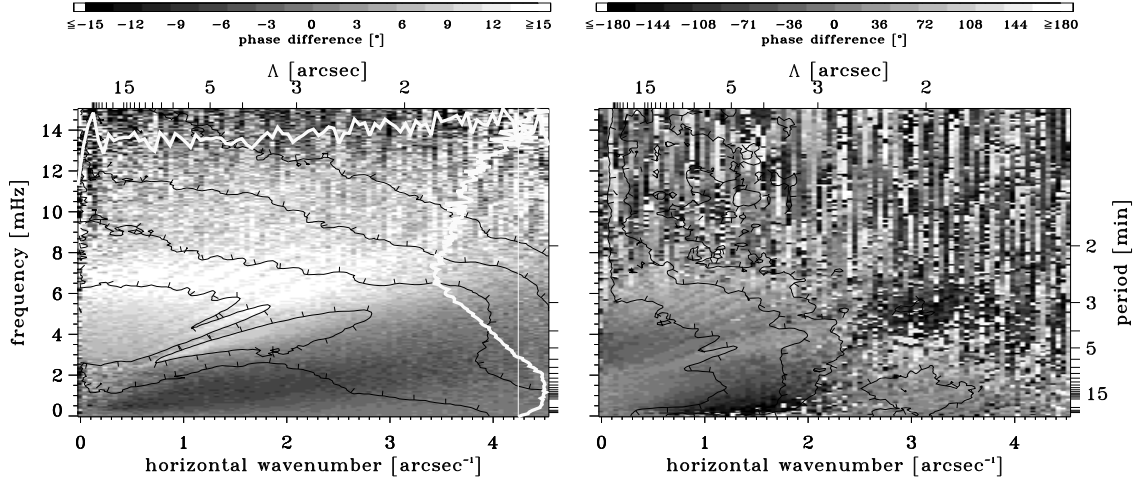}}
  \caption[]{
   Two-dimensional Fourier spectra from TRACE.  Lefthand graph:
   $\Delta \phi(1700\!-\!1600)$ intensity phase difference.  Axes:
   horizontal wavenumber $k_h$ and temporal frequency $f$.  The
   corresponding wavelengths and periodicities are specified along the
   top and righthand side.  Greyscale: phase difference coding
   specified in the bar above the graph.  The white curves along the
   sides are the temporal and spatial means, on linear scales in
   arbitrary units.  The contours specify 1700\,--\,1600\,\AA\ coherence
   at values $C=$ 0.4, 0.6, 0.8 and 0.95, with the ticks directed to
   lower values.  Righthand graph: similar $\Delta \phi({\rm
   WL}\!-\!1700)$ phase difference spectrum between white light
   intensity and 1700\,\AA\ intensity.  The contours specify coherence
   levels $C=$ 0.1, 0.2 and 0.5.  Thijs Krijger production.
%RR refs don't work in caption so can't refer to Thijs' thesis
 }
\label{fig:Fourier} 
\end{figure}
%===========================================================================
%% pasted together with cutkf.tex

%%%%%%%%%%%%%%%%%%%%%%%%%%%%%%%%%%%%%%%%%%%%%%%%%%%%%%%%%%%%%%%%%%%%%%%%%%%%
\paragraph{Fourier representations.}
%%%%%%%%%%%%%%%%%%%%%%%%%%%%%%%%%%%%%%%%%%%%%%%%%%%%%%%%%%%%%%%%%%%%%%%%%%%%

Figure~\ref{fig:Fourier} compares the same TRACE data in terms of \kf\
Fourier phase differences, at left between 1700 and 1600\,\AA\
brightness, at right between white-light and 1700\,\AA\ brightness.
The acoustic oscillations (ridges and pseudo-ridges) in a similar
diagram from the May 12, 1998 TRACE data are discussed in Sect.~4.4 of
\citetads{2001A&A...379.1052K}; %T TRACE1
the emphasis here lies on the low-frequency domain.  Note that the
October 14, 1998 TRACE data contain only very weak network (\cf\
Fig.~\ref{fig:canopy}) so that these \kf\ diagrams are dominated by
internetwork behavior.

The lefthand graph shows a prominent wedge-shaped signature of
negative propagation over a large $k_h$ range.  It has high coherence
between 1700 and 1600\,\AA\ modulation.  The wedge location, its
Lamb-line delimitation, and its negative values all suggest
atmospheric gravity waves as cause.  The large coherence implies that
these waves dominate the low-frequency internetwork background in the
ultraviolet at meso-scale wavelengths.

The righthand \kf\ diagram is very noisy but again shows a low-$f$
low-$k$ wedge of negative phase for $k_h < 2$~arcsec$^{-1}$, \ie\ at
mesogranular scales.  It reaches much larger $\Delta \phi$ values due
to the much larger span in formation height and it has much smaller
coherence, but it is qualitatively similar.  At larger $k_h$ the
low-$f$ phase differences flips from large negative to large positive
values at about $k_h = 2.5$~arcsec$^{-1}$, attributed to wraparound in
the $\Delta \phi = [-180,+180]$~deg evaluation.  This makes the
positive blob around $k_h=3$~arcsec$^{-1}$ (outlined by a $C=0.1$
coherence contour) a continuation of the negative dark wedge,
presumably marking reverse granulation.  The slight
counter-correlation signature in the top row of
Fig.~\ref{fig:scatter2} is therefore indeed caused at granular scales.
This high-$k$ contribution is likely to become better defined at
higher angular resolution than TRACE.  The mesoscale negative-phase
blob in the righthand \kf\ diagram describes the non-correlated other
agent, which I attribute to internal gravity waves.

%RR ?? wat is negative phase blob 5 min in UV-WL komega? 

%%%%%%%%%%%%%%%%%%%%%%%%%%%%%%%%%%%%%%%%%%%%%%%%%%%%%%%%%%%%%%%%%%%%%%%%%%%%
\paragraph{Discussion.}
%%%%%%%%%%%%%%%%%%%%%%%%%%%%%%%%%%%%%%%%%%%%%%%%%%%%%%%%%%%%%%%%%%%%%%%%%%%%

In summary, it seems that the slowly-evolving background mesh pattern
in the internetwork parts of chromospheric image sequences is made up
of reversed granulation and internal gravity waves at slightly larger
scales.  

Numerical simulation may address this interpretation beyond
speculation.  Obviously, this should be feasible with realistic
numerical simulations of the solar granulation such as those of
\citetads{1998ApJ...499..914S} %? Nordlund+Stein: I general properties
when equipped with sufficient atmosphere as in
\citetads{2000ApJ...541..468S}. %T Skartlien+Stein+Nordlund: waves collapsar
They should reproduce both granular overshoot and gravity-wave
emission.  The two phenomena are akin but not identical, since
overshoot is a local phenomenon whereas gravity waves spread away from
their source, interfere, and break at relatively large height (\cf\
\citeads{1981ApJ...249..349M}). %T Mihalas+Toomre I

%%%%%%%%%%%%%%%%%%%%%%%%%%%%%%%%%%%%%%%%%%%%%%%%%%%%%%%%%%%%%%%%%%%%%%%%%%%%
\paragraph{Basal flux.}
%%%%%%%%%%%%%%%%%%%%%%%%%%%%%%%%%%%%%%%%%%%%%%%%%%%%%%%%%%%%%%%%%%%%%%%%%%%%

The demonstration in Fig.~\ref{fig:scatter2} that roughly half of the
brightness excess in internetwork grains is contributed by the
low-frequency variation applies likewise to \CaII\ \KtwoV\ grains and
the accompanying \CaIIK\ line-core and inner-wing brightness
modulation.  Since the solar internetwork scene observed in \CaIIK\ is
dominated by the superposition of these acoustic and low-frequency
variations, the corollary is that part of the basal flux observed from
non-active cool stars \citeads[][ and references
therein]{1995A&ARv...6..181S}, %T Schrijver, A&A review
is attributable to similar combination of acoustic and low-frequency
phenomena including gravity waves -- not just acoustics alone.

%%%%%%%%%%%%%%%%%%%%%%%%%%%%%%%%%%%%%%%%%%%%%%%%%%%%%%%%%%%%%%%%%%%%%%%%%%%%
\paragraph{Internetwork fields.}
%%%%%%%%%%%%%%%%%%%%%%%%%%%%%%%%%%%%%%%%%%%%%%%%%%%%%%%%%%%%%%%%%%%%%%%%%%%%

As noted in Section~\ref{sec:overview} the debate on the pistoning of
internetwork grains continues.  It includes pro-and-contra claims on
internetwork fields as grain localiser.  Advocates pro may argue that
the 50\% grain co-localization by the low-frequency chromospheric
background pattern seen in Fig.~\ref{fig:scatter2} results from weak
fields that are swept into intergranular lanes without reaching the
magnetic flux of network elements, as suggested observationally by
\citetads{1999ApJ...514..448L} %T Lin+Rimmele, granular IN fields
and theoretically by the magnetoconvection simulations of
\citetads{2001ApJ...560L.197E}, %? Emonet+Cattaneo: weak field dynamo
on the assumption that such small concentrations share the network
habits of being bright in the chromosphere and displaying
low-frequency variations there.

%RR low-f: CaIIK + Halpha primarily Doppler, not brightness

I rather doubt this alternative explanation because of the
ubiquity and regularity of the background emission, the low
counter-correlation to intergranular lanes for the spatial
``meso''-scales producing the negative-phase blob in
Fig.~\ref{fig:Fourier}, and the issue how intrinsically weak fields
would influence the thermodynamics of the low chromosphere.  The
inverse-amplitude mapping in Fig.~\ref{fig:canopy} discussed below
suggests rather the reverse to me, namely that larger low-frequency
modulation amplitude implies weaker magnetism.  

Nevertheless, internetwork magnetism cannot be rejected as a
chromospheric brightness producer.  I wonder whether it may play a
role in the generation of small-scale reversed granulation, in the
form of kilogauss fluxtubes that are too thin to be magnetographically
observable so far.  One reason for such wondering is the wide extent
of relatively bright areas surrounding chromospheric network
\citeads[the ``intermediate'' pixel class
of][]{2001A&A...379.1052K}, %T TRACE1
suggesting searches for relations between the occurrence of reversed
granulation and magnetic flux density in high-resolution
large-sensitivity magnetograms.

%RR gravity waves = Zhugzhda temperature waves?

%% {fig:canopy}
%===========================================================================
\begin{figure}
   \centerline{\includegraphics[width=\textwidth]{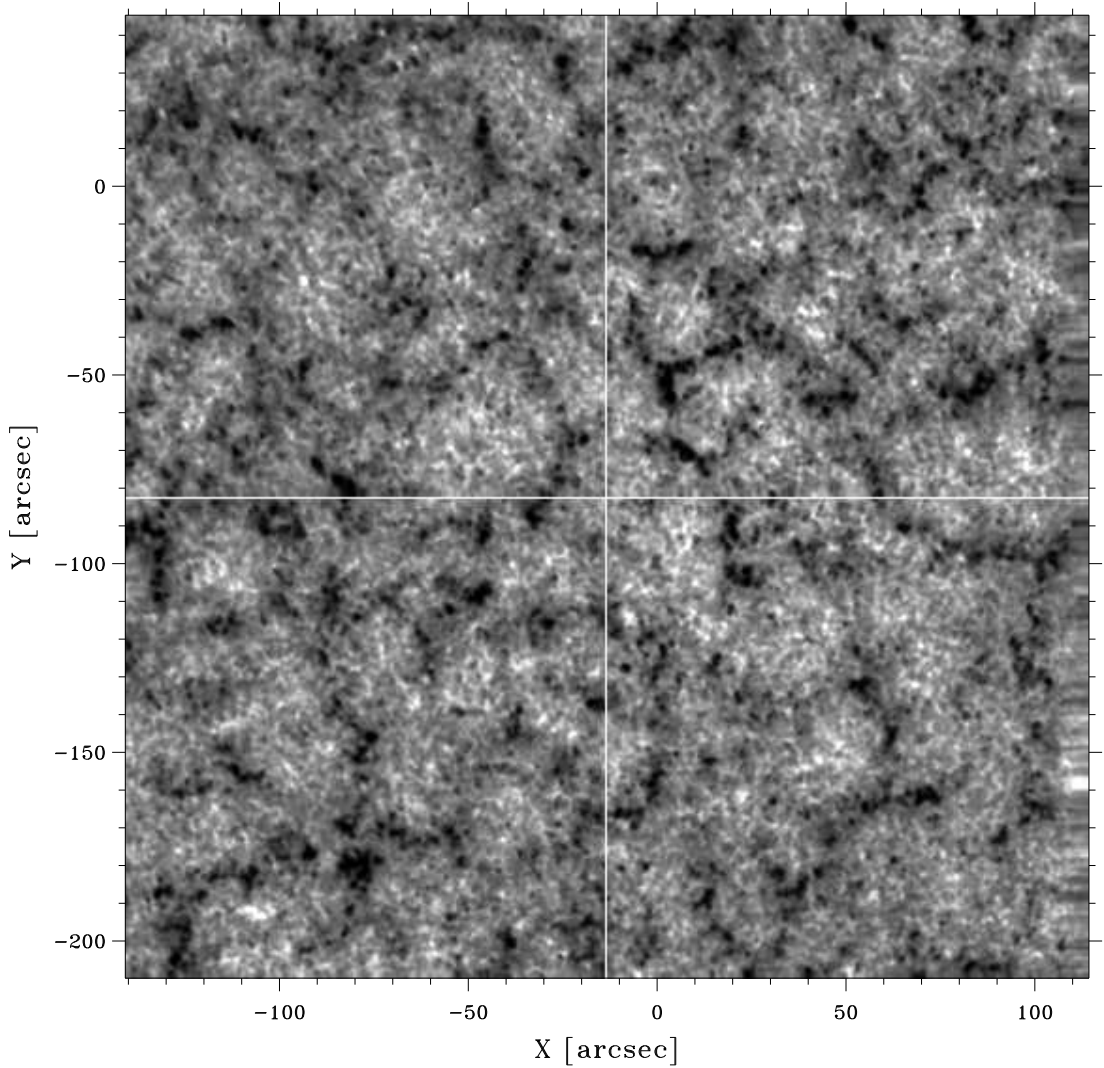}}
   \caption[]{ Temporally-averaged 1700\,\AA\ brightness on an
     inverted greyscale.  The white lines mark division into smaller
     quadrants for processing.  The yet smaller subfield of
     Fig.~\ref{fig:imslices} is near the lower-left corner. The
     horizontal striping near the right edge results from solar
     rotation correction.  }
\label{fig:canopy} 
\end{figure}
%===========================================================================
%RR inverted = 1/x, not reversed = 1-x

%%%%%%%%%%%%%%%%%%%%%%%%%%%%%%%%%%%%%%%%%%%%%%%%%%%%%%%%%%%%%%%%%%%%%%%%%%%%
\paragraph{Inverse canopy mapping.}
%%%%%%%%%%%%%%%%%%%%%%%%%%%%%%%%%%%%%%%%%%%%%%%%%%%%%%%%%%%%%%%%%%%%%%%%%%%%

Finally, Fig.~\ref{fig:canopy} serves to illustrate a speculation
bringing me back to the first item in my review above: the definition
of the canopy as lower chromosphere boundary.  It shows a one-hour
temporal average of the 1700\,\AA\ brightness over the full TRACE
field, with inverted (1/value) greyscale.  The inversion de-emphasizes
network grains by making them dark.  The hour-long averaging reduces
the three-minute component; remaining enhancements from internetwork
grain nonlinearity (non-sinusoidal spikes) also become dark.
 %RR you should average T_b instead, feller
Thus, the brightest features in this display mark locations where the
low-frequency background modulation reaches its deepest minima without
contamination by other phenomena.

Figure~\ref{fig:canopy} shows a grainy pattern.  The brightest grains
mark preponderance of extreme low-frequency minima.  In
reversed-granulation terms, these should correspond to long-lived or
repetitive bright granules surviving a full hour of temporal
averaging.
%RR they do (DOT)
Their distribution over the many network cells covered by
Fig.~\ref{fig:canopy} clearly shows preference for the internetwork
centers, with striking avoidance of the cell boundaries.

My speculation is that such preferential occurrence of low-frequency
extrema marks locations where the magnetic canopy reaches larger
height than elsewhere.  In the gravity-wave interpretation, amplitude
reduction may occur where the field becomes strong enough to convert
gravity waves into other modes or to reflect them.  The areas
containing the brightest inverse grains in Fig.~\ref{fig:canopy} would
then have the highest canopy, giving the waves more room to increase
their amplitude -- I suspect up to wave-breaking height.

%RR but they move slantedly.  But 500 km up is still within big granule.
%RR in DOT the correlation should then increase with smoothing (it does)
%RR what is the breaking height?
%RR big granule may also generate dark hole in spreading gravwavs

A reverse speculation by our meeting's co-director R.~Erdelyi is that
magnetic resonant absorption enhances specific wave modes at the
canopy height and that large apparent amplitude may mark such
resonance (\cf\
\citeads{1997A&A...326.1241C}; %? Cades+Sik+ERdelyi+Goossens: res absorpt
\citeads{2001A&A...372L..17P}). %? Pinter+Erdelyi+New: p-mode conversion

%%%%%%%%%%%%%%%%%%%%%%%%%%%%%%%%%%%%%%%%%%%%%%%%%%%%%%%%%%%%%%%%%%%%%%%%%%%%
\section{Conclusion}  \label{sec:conclusion}
%%%%%%%%%%%%%%%%%%%%%%%%%%%%%%%%%%%%%%%%%%%%%%%%%%%%%%%%%%%%%%%%%%%%%%%%%%%%

Both the review in Section~\ref{sec:overview} and the ultraviolet
background analysis in Section~\ref{sec:background} are
observationally oriented.  However, the major advance in chromospheric
wave research in the past years is, in my view, the advent of
realistic detailed simulations, in particular radiation
(magneto-)\,hydrodynamic simulations casting Scandinavians and Robert
F. Stein as coauthors.  Of course, I allude to the
\citetads{1997ApJ...481..500C} %C Carlsson+Stein, H2v grain simulation
acoustic propagation simulation, the
\citetads{2000ApJ...541..468S} %T Skartlien+Stein+Nordlund, waves collapsar
acoustic excitation simulation, and the
\citetads{2002ApJ...564..508R} %T Rosenthal++, IN reflections
shaken-fluxtube and canopy-interaction MHD simulation.  An important
strength of these simulations is that they are unusually close to
observations, firstly in aiming to reproduce particular observed
phenomena rather than prove grand presupposed concepts, secondly in
the use of elaborate observation-like diagnostics to analyse
simulation physics rather than just showing overall results.

The issues discussed here suggest the following simulation to-do
items: \leftmargini=5ex
\begin{itemize} \vspace*{-1ex} \itemsep=0ex
   \item shaken-fluxtube \CaII\ \HK\ emission;
   \item cool-star Wilson-Bappu relation;
   \item umbral flashes and running penumbral waves;
   \item granular overshoot and gravity wave excitation;
   \item gravity wave interaction with fluxtubes and canopies;
   \item cool-star basal flux evaluation.\\
\end{itemize}

%%%%%%%%%%%%%%%%%%%%%%%%%%%%%%%%%%%%%%%%%%%%%%%%%%%%%%%%%%%%%%%%%%%%%%%%%%%%

{\small \noindent {\bf Acknowledgements.}  I thank the organizers for
  inviting me to a very good workshop and for leniency in their
  editorial deadline coercion.
% \footnote{Note added December 5, 2010: 
%     I became less happy with their lack of diligence in forwarding the
%     content of their proceedings to ADS where this paper is not even
%     listed.}.  
I am indebted to Thijs Krijger for four years of
  pleasant collaboration towards his PhD, for reducing and
  Fourier-analyzing the TRACE data, and for producing
  Fig.~\ref{fig:Fourier}.  I am also indebted to T.D.~Tarbell for
  programming and scheduling TRACE, and to T.J.~Bogdan, B.~Fleck,
  S.S.~Hasan, O.V.~Khomenko, R.I.~Kostik, N.G.~Shchukina, G.~Severino,
  R.F.~Stein and Th.~Straus for inspiring discussions, some held
  within the collaborative framework of and funded by NATO grant
  PST.CLG.97501, INTAS grant 00-00084, and the EC--TMR European Solar
  Magnetometry Network.}

 %%%%%%%%%%%%%%%%%%%%%%%%%%%%%%%%%%%%%%%%%%%%%%%%%%%%%%%%%%%%%%%%% REFERENCES
{\small \bibliographystyle{aa}
\bibsep=0ex \bibhang=3ex
\bibliography{journals,/tmp/rjrfiles.bib,/tmp/adsfiles.bib}
}
 
\end{document}